\documentclass{ifacconf}

\usepackage{graphicx}      
\usepackage{natbib}        
\usepackage{graphics} 
\usepackage{amsmath} 
\usepackage{amssymb}  
\usepackage{bm}
\usepackage{mathtools}
\usepackage{color}
\usepackage{algpseudocode}
\usepackage{algorithm}

\usepackage{amsmath}

\begin{document}
\begin{frontmatter}

\title{Towards targeted exploration for non-stochastic disturbances\thanksref{footnoteinfo}} 

\thanks[footnoteinfo]{F. Allg\"ower is thankful that his work was funded by Deutsche Forschungsgemeinschaft (DFG, German Research Foundation) under Germany’s Excellence Strategy - EXC 2075 - 390740016 and under grant 468094890. J. K\"ohler acknowledges the support of the Swiss National Science Foundation under the NCCR Automation (grant agreement 51NF40\_180545) and an ETH Career Seed Award funded through the ETH Zurich Foundation. J. Venkatasubramanian thanks the International Max Planck Research School for Intelligent Systems (IMPRS-IS) for supporting her.}

\author[First]{Janani Venkatasubramanian} 
\author[Second]{Johannes K\"ohler} 
\author[Third]{Mark Cannon}
\author[First]{Frank Allg\"ower}

\address[First]{Institute for Systems Theory and Automatic Control, University of Stuttgart, 70550 Stuttgart, Germany. (email:\{janani.venkatasubramanian, frank.allgower\}@ist.uni-stuttgart.de)}
\address[Second]{Institute for Dynamic Systems and Control, ETH Z\"urich, Z\"urich CH-80092, Switzerland. (email: jkoehle@ethz.ch)}
\address[Third]{Control Group, University of Oxford, Parks Road, Oxford OX1 3PJ, United Kingdom.(e-mail: mark.cannon@eng.ox.ac.uk)}

\begin{abstract}
We present a novel targeted exploration strategy for linear time-invariant systems without stochastic assumptions on the noise, i.e., without requiring independence or zero mean, allowing for deterministic model misspecifications. This work utilizes classical data-dependent uncertainty bounds on the least-squares parameter estimates in the presence of energy-bounded noise. We provide a sufficient condition on the exploration data that ensures a desired error bound on the estimated parameter. Using common approximations, we derive a semidefinite program to compute the optimal sinusoidal input excitation. Finally, we highlight the differences and commonalities between the developed non-stochastic targeted exploration strategy and conventional exploration strategies based on classical identification bounds through a numerical example.
\end{abstract}

\begin{keyword}
 Experiment Design, Identification for Control, Uncertainty Quantification
\end{keyword}

\end{frontmatter}


\section{Introduction}

Estimation and control design require an accurate model in order to infer unknown parameters and satisfactorily regulate a system's behaviour. The system model can be identified based on data gathered in an experiment. \textit{Targeted exploration} and \textit{optimal experiment design} \citep{pronzato2008optimal, gevers1986optimal} enhance control design by strategically gathering informative data that aids in deriving accurate models. In particular, targeted exploration inputs are designed in such a way that the consequent reduction in model uncertainty ensures: $\mathrm{(i)}$ the identified model meets a desired accuracy, and/or $\mathrm{(ii)}$ the achievement of a desired control goal and performance objective, see e.g., \citep{jansson2005input, bombois2006least,bombois2021robust,barenthin2008identification, larsson2016application, umenberger2019robust, ferizbegovic2019learning, venkatasubramanian2020robust, venkatasubramanian2023sequential}.

The prototypical problem focuses on targeted exploration in order to identify model parameters with a desired error bound \citep{jansson2005input, bombois2021robust, umenberger2019robust}. Most works in this direction assume that the process noise is stochastic, e.g., normally distributed with known mean and variance. Such disturbance models can be utilized to provide high-probability credibility regions for the parameters that can be approximately predicted and optimized \citep{umenberger2019robust}. However, unmodeled dynamics or non-linearities result in additional deterministic model mismatch, which cannot be explained by independent stochastic noise~\citep{sarker2023accurate}. Identification results that account for deterministic noise can be found in~\citep{fogel1979system, sarker2023accurate, bisoffi2021trade, van2020noisy}.



In the proposed approach, we utilize the data-dependent non-stochastic uncertainty bound, derived by \cite{fogel1979system}, to design a targeted exploration strategy that ensures a desired error bound on the estimated parameters. As one of our primary contributions, we derive a sufficient condition on the exploration data, based on the data-dependent uncertainty bound that ensures the desired closeness to the true parameters. We consider multisine exploration inputs of specific frequencies and optimized amplitudes. Utilizing common approximations  \citep{larsson2016application} yields a semidefinite program-based design that allows us to compute an exploration strategy with minimal energy. Finally, through a numerical example, we show that a different distribution of energy over the amplitudes, corresponding to the different frequencies of the multisine inputs, is optimal in the case of energy-bounded disturbances, in contrast to stochastic disturbances. We demonstrate that the designed exploration strategy for non-stochastic noise yields parameter estimates with lower error in the presence of unmodeled system nonlinearities that are energy-bounded, in comparison to stochastic exploration strategies designed for classical stochastic noise.

\section{Problem Statement}
\subsubsection*{Notation:} The transpose of a matrix $A \in \mathbb{R}^{n \times m}$ is denoted by $A^\top$. The positive definiteness of a matrix $A \in \mathbb{R}^{n \times n}$ is denoted by $A = A^\top \succ 0$. The Kronecker product operator is denoted by $\otimes$. The operator $\mathrm{vec}(\cdot)$ stacks the columns of a matrix to form a vector. The operator $\textnormal{diag}(A_1,\dots,A_n)$ creates a block diagonal matrix by aligning the matrices $A_1,\dots,A_n$ along the diagonal starting with $A_1$ in the upper left corner. The Euclidean norm and weighted Euclidean norm for a vector $x\in\mathbb{R}^n$ and a matrix $P \succ 0$ are denoted by $\|x\|=\sqrt{x^\top x}$ and $\|x\|_P=\sqrt{x^\top P x}$, respectively.


\subsection{Setting}
Consider a discrete-time, linear time-invariant system of the form
\begin{align}\label{eq:sys}
x_{k+1}=A_\mathrm{tr} x_k + B_\mathrm{tr} u_k + w_k,
\end{align}
 where $x_k \in \mathbb{R}^{n_\mathrm{x}}$ is the state, $u_k \in \mathbb{R}^{n_\mathrm{u}}$ is the control input, and $w_k \in \mathbb{R}^{n_\mathrm{x}}$ is the disturbance. In our setting, we assume that state is directly measurable. Furthermore, instead of assuming that the disturbance follows some stochastic distribution, we assume that it is energy-bounded.
 \begin{assum}\label{a0} The disturbance $w$ is energy-bounded, i.e., there exists a known constant $\gamma_\mathrm{w}> 0$ such that 
 \begin{align}\label{eq:noisebound}
     \sum_{i=0}^{T-1}\|w_k\|^2 \leq \gamma_\mathrm{w}.
 \end{align}
 \end{assum}
 

\subsubsection*{Exploration goal:} The true system parameters $\theta_\mathrm{tr} =\text{vec}[A_\mathrm{tr}$, $B_\mathrm{tr}]$ are not precisely known. Hence, exploratory inputs should be designed to excite the system to gather informative data. Specifically, our objective is to design exploration inputs that excite the system in a manner as to obtain an estimate $\hat{\theta}_T=\text{vec}[\hat{A}_T,\hat{B}_T]$ that satisfies 
\begin{align}\label{eq:exp_goal2}
    (\theta_\mathrm{tr}-\hat{\theta}_T)^\top D_\mathrm{des} (\theta_\mathrm{tr}-\hat{\theta}_T)\leq 1,
\end{align}
where $D_\mathrm{des} \succ 0$ is a user-defined matrix characterizing closeness of $\hat{\theta}_T$ to $\theta_\mathrm{tr}$.

In order to achieve this goal, we first present a bound on parameter estimation from time-series data for energy-bounded disturbances in Section \ref{sec:prelim}. As the main result, we derive a sufficient condition on the exploration data based on the data-dependent uncertainty bound in order to achieve the exploration goal in Section \ref{sec:exploration}. We later provide tractable approximations and obtain an SDP that can be solved to yield exploration inputs. Finally, in Section \ref{sec:example}, we compare the derived exploration strategy based on the non-stochastic data dependent uncertainty bound with an exploration strategy based on a classical stochastic uncertainty bound.

\section{Uncertainty bound}\label{sec:prelim}
In this section, we discuss a data-dependent uncertainty bound on the parameter estimates in the presence of energy-bounded disturbances \citep{fogel1979system}. Given observed data $\mathcal{D}_T=\{x_k, u_k\}_{k=0}^{T-1}$ of length $T \in \mathbb{N}$, the objective is to quantify the uncertainty associated with the unknown parameters $\theta_\mathrm{tr}$. Henceforth, we denote $\phi_k=[x_k^\top\, u_k^\top]^\top \in \mathbb{R}^{n_\phi}$ where $n_\phi=n_\mathrm{x}+n_\mathrm{u}$. The system \eqref{eq:sys} can be re-written in terms of parameter $\theta_\mathrm{tr}$ as
 \begin{align}
 x_{k+1}=(\phi_k^\top \otimes I_{n_\mathrm{x}})\theta_\mathrm{tr} + w_k.
 \end{align}
From the standard least squares formulation, we obtain the expressions for the mean $\hat{\theta}_T=\textnormal{vec}[\hat{A}_T,\hat{B}_T]$ and covariance $P_\theta$ as
\begin{align}\label{eq:mean_est}
    \hat{\theta}_T=P_{\theta}\sum_{k=0}^{T-1} (\phi_k^\top \otimes I_{n_\mathrm{x}})^\top x_{k+1}
\end{align}
and
\begin{align}\label{eq:covar_est}
    P_{\theta}^{-1}=\left( \sum_{k=0}^{T-1} \phi_k \phi_k^\top \right) \otimes I_{n_\mathrm{x}}.
\end{align}
The following lemma provides a non-falsified region for the uncertain parameters $\theta$.

\begin{lem}\label{lem:thetat} \citep{fogel1979system} Given data set $\mathcal{D}_T$ and disturbance with energy bound $\gamma_\mathrm{w}$ (cf. Assumption \ref{a0}), the set of non-falsified parameters $\theta$ is given by
\begin{align}\label{eq:Theta_T}
    \mathbf{\Theta}:=\left\{ \theta: (\theta-\hat{\theta}_T )^\top{P_{\theta}^{-1}} (\theta-\hat{\theta}_T) \leq G \right\}
\end{align}
where
\begin{align}\label{eq:G}
    G :=\gamma_\mathrm{w} + \|\hat{\theta}_T\|_{P_{\theta}^{-1}}^2-\sum_{k=0}^{T-1}\|x_{k+1}\|^2.
\end{align}
\end{lem}
\begin{pf}
The energy constraint on the disturbance in \eqref{eq:noisebound} yields the following non-falsified set:
\begin{align}\label{eq:data_thetat}
\mathbf{\Theta}&=\left\{ \theta: \sum_{k=0}^{T-1} \|x_{k+1}-(\phi_k^\top \otimes I_{n_\mathrm{x}})\theta \|^2 \leq \gamma_\mathrm{w} \right\}
\end{align}
which can be equivalently written as
\begin{align}
\nonumber
    \theta^\top \underbrace{\left( \sum_{k=0}^{T-1} (\phi_k \otimes I_{n_\mathrm{x}}) (\phi_k^\top \otimes I_{n_\mathrm{x}}) \right)}_{=:P_{\theta}^{-1}}\theta + \sum_{k=0}^{T-1}\|x_{k+1}\|^2\\
    -2\left(\sum_{k=0}^{T-1} x_{k+1}^\top (\phi_k^\top \otimes I_{n_\mathrm{x}}) \right)\theta \leq \gamma_\mathrm{w}.
\end{align}

By using \eqref{eq:mean_est} and \eqref{eq:covar_est}, and by completing the squares, we get
\begin{align}\label{eq:Gtbound}
    \|\theta-\hat{\theta}_T\|_{P_{\theta}^{-1}}^2 \leq \gamma_\mathrm{w} + \|\hat{\theta}_T\|_{P_{\theta}^{-1}}^2-\sum_{k=0}^{T-1}\|x_{k+1}\|^2 =G,
\end{align}
which is equivalent to \eqref{eq:Theta_T}. $\hfill\square$
\end{pf}

\begin{figure*}[t!]
    \begin{align}\label{eq:schur_pe}
            \begin{bmatrix}
           (\Phi \Phi^\top)\otimes I_{n_\mathrm{x}}-\gamma_\mathrm{w}D_\mathrm{des}+D_\mathrm{des}^{\frac{1}{2}\top}\left((X^\top X) \otimes I_{n_\mathrm{x}n_\phi}\right)D_\mathrm{des}^{\frac{1}{2}}
            &  D_\mathrm{des}^{\frac{1}{2}\top}((X^\top(\Phi^\top \otimes I_{n_\mathrm{x}}) )\otimes I_{n_\mathrm{x}n_\phi})\\
            (((\Phi \otimes I_{n_\mathrm{x}}) X)\otimes I_{n_\mathrm{x}n_\phi})D_\mathrm{des}^{\frac{1}{2}} & ((\Phi \Phi^\top)\otimes I_{n_\mathrm{x}})\otimes I_{n_\mathrm{x}n_\phi}
        \end{bmatrix} \succeq 0.
    \end{align}
\end{figure*}




Notably, the non-falsified set $\bm{\Theta}$ provides an exact characterization of the set of parameters explaining the data, given Assumption \ref{a0}. The ellipsoid \eqref{eq:Theta_T} obtained from energy-bounded constraints is characterized by a vector $\hat{\theta}_T$ and a matrix $P_\theta$, which coincide with the mean and covariance of the least squares estimator for linear systems with Gaussian noise \citep[Prop. 2.1]{umenberger2019robust}. However, a crucial difference between the robust case of energy-bounded disturbance and the stochastic Gaussian case is that the scaling $G$ \eqref{eq:G} of the bounding ellipsoid is data-dependent. Similar non-falsified sets are considered in \citep{van2020noisy,berberich2022combining}. In the next section. we derive a targeted exploration strategy using the data-dependent uncertainty bound in Lemma \ref{lem:thetat}.

\section{Targeted Exploration}\label{sec:exploration}
In this section, we propose a targeted exploration strategy based on the robust uncertainty bound on the data obtained through the process of exploration derived in Lemma \ref{lem:thetat}. There exists a large body of literature on targeted exploration strategies that assume independent and identically distributed (i.i.d.) stochastic disturbances \citep{ larsson2016application, bombois2021robust, venkatasubramanian2023sequential}. In contrast, we propose an exploration strategy based on the data-dependent uncertainty bound for energy-bounded disturbances derived in Lemma \ref{lem:thetat}. In particular, we first derive sufficient conditions on the exploration data to achieve the exploration goal \eqref{eq:exp_goal2} (cf. Section \ref{subsec:sufficient}). Since this condition is non-convex in the decision variables, a convex relaxation procedure is carried out (cf. Section \ref{subsec:convexrel}). Furthermore, we provide tractable approximations in Section \ref{subsec:tractable}, and obtain an SDP that can be solved to yield exploration inputs.

\subsection{Exploration strategy}
The exploration input sequence takes the form
\begin{align}\label{eq:exploration_controller}
u_k=\sum_{i=1}^{L} a_i \cos(2 \pi \omega_i k), \quad\, k=0,\dots,T-1
\end{align}
where $T$ is the exploration time and $a_i \in \mathbb{R}^{n_\mathrm{u} \times 1}$ are the amplitudes of the sinusoidal inputs at $L$ distinctly selected frequencies $\omega_i \in \Omega_T :=\{0,1/T,...,(T-1)/T\}$.
We denote $U_\mathrm{e}=\mathrm{diag}(a_1,\dots,a_{L}) \in \mathbb{R}^{Ln_\mathrm{u} \times L}$. The exploration input is computed such that it excites the system sufficiently with minimal control energy, based on the initial parameter estimates. To this end, we require that the control energy at each time instant does not exceed $\gamma^2_\mathrm{e}$, i.e., $\sum_{i=1}^L \|a_i\|^2= \mathbf{1}_{L}^\top U_\mathrm{e}^\top U_\mathrm{e} \mathbf{1}_{L} \preceq \gamma_\mathrm{e}^2$ where $\mathbf{1}_{L} \in \mathbb{R}^{L\times 1}$ is a vector of ones, and the bound $\gamma_\mathrm{e} \geq 0$ is desired to be small. Using the Schur complement, this criterion is equivalent to 
\begin{align}\label{eq:min_energy_cost}
S_{\textnormal{energy-bound}}(\gamma_\mathrm{e},U_\mathrm{e})\coloneqq \begin{bmatrix} 
\gamma_\mathrm{e} & \mathbf{1}_{L}^\top U_\mathrm{e}^\top \\ U_\mathrm{e} \mathbf{1}_{L} & \gamma_\mathrm{e} I
\end{bmatrix} \succeq 0.
\end{align}

\begin{rem} \label{rem:2} Since we consider only open-loop inputs in our exploration strategy, we require $A_\mathrm{tr}$ to be Schur stable. An exploration input of the form in \eqref{eq:exploration_controller} with an additional linear feedback, i.e., $v_k=u_k+Kx_k$, which robustly stabilizes the initial estimate and prior uncertainty, may be considered if it is not known whether the system is Schur stable.
\end{rem}

\subsection{Sufficient conditions for targeted exploration}\label{subsec:sufficient}
Given the data-dependent uncertainty bound in Lemma~\ref{lem:thetat}, in what follows, we determine the conditions that the exploration data has to satisfy to achieve the exploration goal. In order to simplify the exposition, we denote
\begin{align}\label{eq:phi_compact}
\nonumber
    \Phi&=[\phi_0,\dots,\phi_{T-1}] \in \mathbb{R}^{n_\phi\times T},\\
    X^\top&=[x_1^\top,\dots, x_T^\top] \in \mathbb{R}^{1 \times Tn_\mathrm{x}}.
\end{align}
Then, the mean estimate \eqref{eq:mean_est} is given by
\begin{align}\label{eq:mean_vec}
    \hat{\theta}_T = P_{\theta} (\Phi \otimes I_{n_\mathrm{x}}) X
\end{align}
and the covariance $P_{\theta}$ \eqref{eq:covar_est} can be computed from
\begin{align}\label{eq:invP}
P_{\theta}^{-1}=(\Phi \Phi^\top) \otimes I_{n_\mathrm{x}}.
\end{align}

We denote the Cholesky decomposition of $D_\mathrm{des}$ as $D_\mathrm{des}=D_\mathrm{des}^{\frac{1}{2}\top}D_\mathrm{des}^{\frac{1}{2}}$ where $D_\mathrm{des}^{\frac{1}{2}}$ is an upper triangular matrix. The following theorem presents a sufficient condition to ensure that the exploration goal is achieved.


\begin{thm}\label{prop:lowerdt}
Let Assumption \ref{a0} hold. Suppose $\Phi$ and $X$ satisfy
\begin{align}\label{eq:pe5}
\nonumber
    \begin{bmatrix}
        (\Phi \Phi^\top)\otimes I_{n_\mathrm{x}} - \gamma_\mathrm{w} D_\mathrm{des}& 0\\
        0 & 0
    \end{bmatrix}&\\ 
    +\underbrace{\begin{bmatrix}
       D_\mathrm{des}^{\frac{1}{2}\top}(X^\top \otimes I_{n_\mathrm{x}n_\phi}) \\ (\Phi \otimes I_{n_\mathrm{x}}) \otimes I_{n_\mathrm{x}n_\phi}
    \end{bmatrix}}_{Y} \begin{bmatrix}
       D_\mathrm{des}^{\frac{1}{2}\top}(X^\top \otimes I_{n_\mathrm{x}n_\phi}) \\ (\Phi \otimes I_{n_\mathrm{x}}) \otimes I_{n_\mathrm{x}n_\phi}
    \end{bmatrix}^\top &\succeq 0.
\end{align}
Then, the estimate $\hat{\theta}_T$ computed as in \eqref{eq:mean_est} satisfies the exploration goal of $\theta_\mathrm{tr} \in \bm{\hat{\Theta}}_T$ \eqref{eq:exp_goal2}.
\end{thm}

\begin{pf}
The bound in \eqref{eq:Gtbound} can be re-written as
\begin{align}\label{eq:Gtbound_re}
\nonumber
    &(\theta-\hat{\theta}_T)^\top ((\Phi \Phi^\top) \otimes I_{n_\mathrm{x}}) (\theta-\hat{\theta}_T)\\
\nonumber
    \overset{\eqref{eq:mean_vec}}{\leq} & \gamma_\mathrm{w}-X^\top X +X^\top(\Phi^\top\otimes I_{n_\mathrm{x}}) P_{\theta,T} (\Phi\otimes I_{n_\mathrm{x}})X\\
    \overset{\eqref{eq:invP}}{=} & \gamma_\mathrm{w}-X^\top X +X^\top ((\Phi^\top (\Phi \Phi^\top)^{-1}\Phi) \otimes I_{n_\mathrm{x}} )X.
\end{align}

By applying the Schur complement twice to \eqref{eq:Gtbound_re}, we get
\begin{align}\label{eq:bound_schur}
\nonumber
    &(\theta-\hat{\theta}_T) (\theta-\hat{\theta}_T)^\top\\ 
    \nonumber \preceq & (\gamma_\mathrm{w}-X^\top X +X^\top ((\Phi^\top (\Phi \Phi^\top)^{-1}\Phi) \otimes I_{n_\mathrm{x}} )X) \\& \cdot \left((\Phi \Phi^\top)^{-1}\otimes I_{n_\mathrm{x}}\right).
\end{align}
Inequality \eqref{eq:pe5} can be written as \eqref{eq:schur_pe}. By applying the Schur complement to \eqref{eq:schur_pe}, we get 
\begin{align}
\nonumber&(\Phi \Phi^\top)\otimes I_{n_\mathrm{x}} - \gamma_\mathrm{w}D_\mathrm{des} + D_\mathrm{des}^{\frac{1}{2}\top}\left((X^\top X)\otimes I_{n_\mathrm{x}n_\phi}\right)D_\mathrm{des}^{\frac{1}{2}} \\ 
\nonumber
    &-D_\mathrm{des}^{\frac{1}{2}\top}\left((X^\top ((\Phi^\top (\Phi \Phi^\top)^{-1}\Phi) \otimes I_{n_\mathrm{x}}  ) X ) \otimes I_{n_\mathrm{x}n_\phi} \right)D_\mathrm{des}^{\frac{1}{2}}\\
\label{eq:pe2}
    \succeq &\;0
\end{align}
which can be written as
\begin{align}\label{eq:pe_ineq}
\nonumber
    (\Phi \Phi^\top)\otimes I_{n_\mathrm{x}} \succeq & (\gamma_\mathrm{w}-X^\top X \\
    &+X^\top ((\Phi^\top (\Phi \Phi^\top)^{-1}\Phi) \otimes I_{n_\mathrm{x}} )X) D_\mathrm{des}
\end{align}
since the term $\gamma_\mathrm{w}-X^\top X+X^\top ((\Phi^\top (\Phi \Phi^\top)^{-1}\Phi) \otimes I_{n_\mathrm{x}} )X$ is scalar.
Furthermore, by inserting \eqref{eq:pe_ineq} in \eqref{eq:bound_schur}, we get 
\begin{align}\label{eq:expgoal_schur}
    (\theta-\hat{\theta}_T) (\theta-\hat{\theta}_T)^\top \preceq D_\mathrm{des}^{-1}.
\end{align}
Finally, applying the Schur complement twice to \eqref{eq:expgoal_schur} yields the exploration goal \eqref{eq:exp_goal2}. $\hfill\square$
\end{pf}

Inequality \eqref{eq:pe5} uses $X$ and $\Phi$, which depend on the amplitudes of the sinusoidal signals $U_\mathrm{e}$ \eqref{eq:exploration_controller}, as well as the disturbance $w$. Thus, ensuring the exploration goal \eqref{eq:exp_goal2} results in non-convex constraints in the decision variable $U_\mathrm{e}$. To overcome this problem, we utilize a convex relaxation procedure.

\subsection{Convex relaxation}\label{subsec:convexrel}
The following lemma is utilized to provide sufficient conditions for Inequality \eqref{eq:pe5} that are linear in the decision variable $U_\mathrm{e}$.
\begin{lem}\label{lem:convexrel1} For any matrices $L \in \mathbb{C}^{n \times m}$ and $N \in \mathbb{C}^{n \times m}$, we have:
\begin{align}\label{eq:convex_rel}
    N N^\mathsf{H} \succeq N L^\mathsf{H} + L N^\mathsf{H} - L L^\mathsf{H}.
\end{align}
\end{lem}
\begin{pf} We have
\begin{align*}
N N^\mathsf{H} - N L^\mathsf{H} - L N^\mathsf{H} + L L^\mathsf{H}=(N-IL)(N-IL)^\mathsf{H}\succeq 0
\end{align*}    
and hence, we have \eqref{eq:convex_rel}. $\hfill\square$
\end{pf}


The following proposition provides a sufficient condition that is linear in $\Phi$ and $Y$ which, if satisfied, ensures the exploration goal \eqref{eq:exp_goal2}.

\begin{prop} Suppose there exist matrices $\Phi$, $Y$, $L_1 \in \mathbb{R}^{n_\phi \times T}$ and $L_2 \in \mathbb{R}^{(n_\mathrm{x}n_\phi+n_\mathrm{x}^2 n_\phi^2) \times T n_\mathrm{x}^2 n_\phi}$ that satisfy the following matrix inequality:
\begin{align}\label{eq:pe6}
\nonumber
&S_\textnormal{exploration}(\Phi,Y,L_1,L_2,\gamma_\mathrm{w}, D_\mathrm{des})\\
\nonumber
:=&\begin{bmatrix}
        (\Phi L_1^\top + L_1\Phi^\top -L_1 L_1^\top)\otimes I_{n_\mathrm{x}} - \gamma_\mathrm{w} D_\mathrm{des}& 0\\
        0 & 0
    \end{bmatrix}\\
\nonumber
    &+Y L_2^\top + L_2 Y^\top - L_2 L_2^\top\\
    \succeq & 0.
\end{align}
Then, the estimate $\hat{\theta}_T$ computed as in \eqref{eq:mean_est} satisfies the exploration goal \eqref{eq:exp_goal2}.
\end{prop}

\begin{pf} Starting from Inequality \eqref{eq:pe6}, we have
\begin{align*}
    0 \preceq &  \begin{bmatrix}
        (\Phi L_1^\top + L_1\Phi^\top -L_1 L_1^\top)\otimes I_{n_\mathrm{x}} - \gamma_\mathrm{w} D_\mathrm{des}& 0\\
        0 & 0
    \end{bmatrix}\\
    &+Y L_2^\top + L_2 Y^\top - L_2 L_2^\top\\
    \overset{\eqref{eq:convex_rel}}{\preceq}&\begin{bmatrix}
        (\Phi \Phi^\top)\otimes I_{n_\mathrm{x}} - \gamma_\mathrm{w} D_\mathrm{des}& 0\\
        0 & 0
    \end{bmatrix}\\
    &+Y L_2^\top + L_2 Y^\top - L_2 L_2^\top\\
    \overset{\eqref{eq:convex_rel}}{\preceq}&\begin{bmatrix}
        (\Phi \Phi^\top)\otimes I_{n_\mathrm{x}} - \gamma_\mathrm{w} D_\mathrm{des}& 0\\
        0 & 0
    \end{bmatrix}+Y Y^\top.
\end{align*}
 Hence, if there exists a matrices $\Phi$ and $Y$ satisfying \eqref{eq:pe6}, then the condition in Theorem \ref{prop:lowerdt} is satisfied and the exploration goal \eqref{eq:exp_goal2} is achieved. $\hfill\square$
\end{pf}
The bound \eqref{eq:convex_rel} derived in Lemma \ref{lem:convexrel1} is tight if $L_1=\Phi$ and $L_2=Y$. Inequality \eqref{eq:pe6} is linear in $\Phi$ and $Y$ which are composed of $x_k$, $u_k$, $k=0,...,T-1$. Hence, $\Phi$ and $Y$ are linear in the chosen inputs $u_k$ and disturbance $w_k$. However, a few challenges need to be addressed in order to tractably compute the exploration inputs: $\mathrm{(i)}$ since the true dynamics $A_\mathrm{tr},\,B_\mathrm{tr}$ are unknown, the linear mapping from the input sequence to the state sequence is not known, and $\mathrm{(ii)}$ the future states $x_k$ depend on the disturbances $w_k$, which are unknown.
Notably, these challenges equally appear in the existing formulations for targeted exploration with stochastic noise assumptions, e.g.,~\citep{barenthin2008identification, larsson2016application}. In the following section, we address these problems using common approximations from the literature and demonstrate how to tractably compute the exploration strategy.

\subsection{Approximations for a tractable solution}\label{subsec:tractable}

In order to tractably compute the exploration strategy, certain approximations and assumptions are made. Although the dynamics $A_\mathrm{tr},\,B_\mathrm{tr}$ are not exactly known, we consider a rough initial estimate of the dynamics $\hat{\theta}_0=\mathrm{vec}([\hat{A}_0,\hat{B}_0])$. Consequently, the challenge is to account for $w$ and the error in $\hat{\theta}_0$ in computing the future states $X,\,\Phi$.

\begin{rem}
     Such an initial estimate may also be inferred from data obtained through a short experiment. In particular, given a data set $\mathcal{D}_0=\{\phi_k \}_{k=-\bar{T}}^{-1}$ obtained from a randomly exciting input, $\hat{\theta}=\mathrm{vec}([\hat{A}_0,\hat{B}_0])$ is the ordinary least squares estimate.
\end{rem}
 

To address this challenge, we consider the following simplifications: $\mathrm{(i)}$ the future states evolve exactly according to $\hat{\theta}_0$, i.e., we neglect the error $\theta_\mathrm{tr}-\hat{\theta}_0$; $\mathrm{(ii)}$ the effect of $w$ on $x$ is small compared to $u$, and hence we neglect this in determining future states (cf., e.g., \cite{larsson2016application}). In what follows, we use the initial estimate to design a targeted exploration strategy. To simplify the exposition, we assume that $n_\mathrm{u}=1$, i.e., $u_k$ is scalar\footnote{For $n_\mathrm{u}>1$, $u_k^{(i)}=\sum_{j=1}^{n_\mathrm{u}}b_j \cos(\omega_i k),\,i=1,\dots,L,$ where $b_j$ is a basis vector comprising zeros, except the $j^{th}$ element, which is 1.}, and $x_0=0$, without loss of generality.
Given the exploration strategy \eqref{eq:exploration_controller}, we obtain a simple/tractable expression for $\Phi$ and $X$ by simulating the open-loop systems $\hat{A}_0,\hat{B}_0$ with $w=0$ and
\begin{align}\label{eq:uki}
    u_k^{(i)}=\cos(2 \pi \omega_i k),\,i=1,\dots,L,
\end{align}
for $k=1,...,T,$ and recording the resulting values in $\hat{X}^{(i)},\,\hat{\Phi}^{(i)}$.
We then obtain
\begin{align}\label{eq:phi_x_hat}
\Phi \approx \hat{\Phi}(U_\mathrm{e})=\sum_{i=1}^L \hat{\Phi}^{(i)}a_i,~ X \approx \hat{X}(U_\mathrm{e})=\sum_{i=1}^L \hat{X}^{(i)}a_i.
\end{align}
which are approximations of $\Phi$ and $X$ linear in $U_\mathrm{e}$.
Further, we determine $\hat{Y}$ linear in $U_\mathrm{e}$ as
\begin{align}
    \hat{Y}(U_\mathrm{e})=\begin{bmatrix}
       D_\mathrm{des}^{\frac{1}{2}\top}(\hat{X}(U_\mathrm{e})^\top \otimes I_{n_\mathrm{x}n_\phi}) \\ (\hat{\Phi}(U_\mathrm{e}) \otimes I_{n_\mathrm{x}}) \otimes I_{n_\mathrm{x}n_\phi}
    \end{bmatrix}.
\end{align}

Hence, Inequality \eqref{eq:pe6} reduces to the following linear matrix inequality in $U_\mathrm{e}$:

\begin{align}\label{eq:exp_LMI}
S_\textnormal{exploration}(\hat{\Phi}(U_\mathrm{e}),\hat{Y}(U_\mathrm{e}),L_1,L_2,\gamma_\mathrm{w}, D_\mathrm{des}) \succeq  0.
\end{align}

As a result, we can pose the exploration problem of achieving the exploration goal \eqref{eq:exp_goal2} with minimal input energy using the following SDP:
\begin{align}
\nonumber
\underset{U_\mathrm{e},\gamma_\mathrm{e}}{\inf}  & \quad \gamma_\mathrm{e} \\
\label{eq:exp_problem}
\text{s.t. }& \quad S_{\textnormal{energy-bound}}(\gamma_\mathrm{e},U_\mathrm{e})\succeq 0\\
\nonumber
& \quad S_\textnormal{exploration}(\hat{\Phi}(U_\mathrm{e}),\hat{Y}(U_\mathrm{e}),L_1,L_2,\gamma_\mathrm{w}, D_\mathrm{des}) \succeq 0.
\end{align}

The optimization variables $U_\mathrm{e}=\textnormal{diag}(a_1,\dots,a_{n_\mathrm{\phi}})$ required for the implementation of the exploration input are obtained from a solution of \eqref{eq:exp_problem}. To eliminate the conservatism stemming from the convex relaxation procedure, we iterate Problem \eqref{eq:exp_problem} multiple times by recomputing $L_1$ and $L_2$ for the next iteration as
\begin{align}\label{eq:L1L2}
    L_1=\tilde{\Phi},\,L_2=\begin{bmatrix}
       \tilde{X}^\top \otimes I_{n_\mathrm{x}n_\phi} \\ (\tilde{\Phi} \otimes I_{n_\mathrm{x}}) \otimes I_{n_\mathrm{x}n_\phi}
    \end{bmatrix},
\end{align}
where $\tilde{\Phi}$ and $\tilde{X}$ are computed as
\begin{align}
\tilde{\Phi}_T=\sum_{i=1}^L \hat{\Phi}_T^{(i)}\tilde{a}_i,\,\tilde{X}_T=\sum_{i=1}^L  \hat{X}_T^{(i)}\tilde{a}_i
\end{align}
where $\tilde{a}_i$ is an arbitrary candidate for the first iteration, and for further iterations, $\tilde{a}_i=a^*_i$ is the solution from the previous iteration. The overall targeted exploration strategy is summarized in Algorithm \ref{alg:main}. The goal of the strategy is to excite the system with exploratory inputs \eqref{eq:exploration_controller} in order to determine model parameters up to a user defined closeness $D_\mathrm{des}$.

\begin{algorithm}[H]
\caption{Non-stochastic targeted exploration}
\label{alg:main}
\begin{algorithmic}[1]
\State Specify exploration length $T$, frequencies $\omega_i,\,i=1,...,L$, energy bound $\gamma_\mathrm{w}$, initial estimates $\hat{A}_0,\,\hat{B}_0$, desired accuracy of parameters $D_\mathrm{des}$.
\State Compute $\hat{\Phi}^{(i)},\,\hat{X}^{(i)}$ using initial estimates $\hat{A}_0,\,\hat{B}_0$ \eqref{eq:phi_x_hat}.
\State Select initial candidate $\tilde{a}_i$, $i=1,...,L$.
\While{$\| \tilde{a}_i - a_i^*\| > \mathrm{tol} $}
\State Compute $L_1,\,L_2$ \eqref{eq:L1L2}.
\State Solve the optimization problem \eqref{eq:exp_problem}.
\State Set $\tilde{a}_i=a^*_i$.
\EndWhile
\State Apply the exploration input \eqref{eq:exploration_controller} for $k=0,...,T-1$.
\State Compute $\hat{\theta}_T$ \eqref{eq:mean_est}; compute parameter set $\bm{\Theta}_T$ \eqref{eq:data_thetat}.
\end{algorithmic}
\end{algorithm}

Note that $S_\textnormal{exploration}(\hat{\Phi}(U_\mathrm{e}),\hat{Y}(U_\mathrm{e}),L_1,L_2,\gamma_\mathrm{w}, D_\mathrm{des})$ \eqref{eq:exp_LMI} is an LMI of dimension  $n_\mathrm{x}n_\phi+(n_\mathrm{x}n_\phi)^2$, which is independent of the exploration time $T$. The time $T$ is only of significance in step 2 of Algorithm \ref{alg:main} while computing the trajectories $\hat{X}^{(i)}$ and $\hat{\Phi}^{(i)}$.

\subsection{Discussion}
The proposed exploration strategy summarized in Algorithm \ref{alg:main} serves as a preliminary bridge between identification methods for systems with non-stochastic disturbances \citep{fogel1979system} with design of targeted exploration \citep{barenthin2008identification}. In case the disturbances in the system is in fact Gaussian, then an energy bound of the form in \eqref{eq:G} still remains valid with high probability. Hence, the non-stochastic uncertainty bound on the parameters remains valid with high probability. On the other hand, in the presence of unmodeled dynamics, which lead to deterministic errors, standard stochastic uncertainty bounds are, in general, incorrect~\citep{sarker2023accurate}. Although the mean $\hat{\theta}_T$ and covariance $P_\theta$ that characterize both the stochastic and non-stochastic parameter uncertainty bounds are the same, there are significant differences: $\mathrm{(i)}$ the scaling $G$ \eqref{eq:G} for the non-stochastic uncertainty bound is data-dependent unlike the stochastic bound, $\mathrm{(ii)}$ the stochastic uncertainty bound is arbitrarily small for arbitrarily small excitation under sufficiently long exploration time $T$, unlike the non-stochastic bound. These differences in uncertainty bound descriptions lead to significant differences in quantifying parameter errors. Therefore, it is necessary to appropriately consider and address these distinctions.





\section{Numerical example}\label{sec:example}
In this section, we consider a simple numerical example and highlight the differences between the developed targeted exploration strategy for non-stochastic disturbances and conventional exploration strategies based on the classical stochastic disturbance assumptions. Numerical simulations were performed on MATLAB using CVX \citep{cvx} in conjunction with the default SDP solver SDPT3. We first introduce the problem setup, and then compare the proposed targeted exploration for non-stochastic disturbances to targeted exploration based on standard stochastic assumptions, e.g., \citep{venkatasubramanian2023sequential}.  We consider a linear system \eqref{eq:sys} taken from \citep{venkatasubramanian2023sequential} with
\begin{align}\label{eq:exsys}
A_\mathrm{tr}=\begin{bmatrix}
0.49 & 0.49 & 0 & 0\\ 0 & 0.49 & 0.49 & 0\\ 0 & 0 & 0.49 & 0.49 \\ 0 & 0 & 0 & 0.49
\end{bmatrix},\;
B_\mathrm{tr}=\begin{bmatrix}
0 \\ 0 \\ 0\\  0.49
\end{bmatrix}.
\end{align}

\subsection{Non-stochastic vs. stochastic targeted exploration}
 \begin{figure}[t]
\begin{center}
\includegraphics[width=0.4\textwidth]{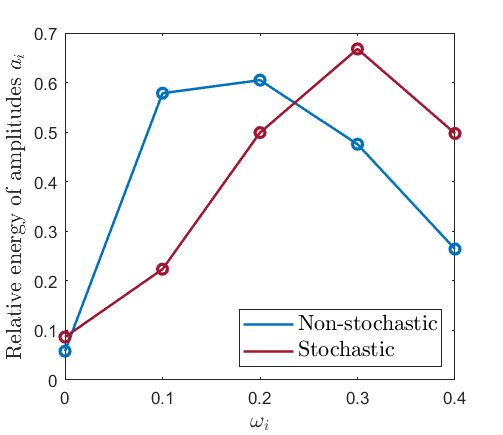}
\end{center}
\caption{Optimal relative energy over different frequencies $\omega_i$ for non-stochastic and stochastic targeted exploration.}
\label{fig:relenergy}
\end{figure}

The goal of targeted exploration is to ensure $\theta_\mathrm{tr}$ is within a user-defined closeness $D_\mathrm{des}$ to the true parameters $\theta_\mathrm{tr}$~\eqref{eq:exp_goal2}. We select $D_\mathrm{des}=I$ and exploration time $T=100$. The initial estimate is given by $\hat{\theta}_0=\theta_\mathrm{tr}+(5 \times 10^{-3})\mathbf{1}_{n_\mathrm{x}n_\phi}$. We select $L=5=n_\phi$ frequencies from $\Omega_{100}$ to yield $\omega_i=\{0,0.1,0.2,0.3,0.4\}$, $i~=~1,...,L$. The solution of Problem~\eqref{eq:exp_problem} provides the corresponding amplitudes $a_i,\,i=1,...,L$. In the context of stochastic targeted exploration, the exploration problem \citep[Prob. (46)]{venkatasubramanian2023sequential} is solved with the same simplifications used to solve the non-stochastic problem. We solve the exploration problem according to Algorithm \ref{alg:main} and the exploration problem based on stochastic assumptions. From Fig. \ref{fig:relenergy}, it can be observed that the optimal relative energy over frequencies $\omega_i$ yield different profiles. This supports our assertion that the optimal exploration depends on the disturbance characterization. Note that the relative energy over frequencies do \textit{not} depend on the absolute value of $\gamma_\mathrm{w}$, $\sigma_\mathrm{w}^2$. 

In order to further asses both the targeted exploration strategies, we compare the estimates obtained after exploration in the presence of unmodeled nonlinearities. Consider a nonlinearity in the model given by
\begin{align}
    w_k=c \begin{bmatrix}
        -\cos (x_{k,1})&0&0&0
    \end{bmatrix}^\top,
\end{align}
where $c=\sqrt{0.1}$ and $x_{k,1}$ is the first element of the state at time $k$. This disturbance is energy-bounded, i.e., Assumption \ref{a0} holds with $\gamma_\mathrm{w}=10$, and $w_k$ is not independent or zero mean. We compute the non-stochastic exploration inputs according to Algorithm \ref{alg:main} and the stochastic targeted exploration inputs according to \citep{venkatasubramanian2023sequential}. For a fair comparison, the stochastic exploration
inputs are scaled such that both strategies apply inputs with the same energy for exploration. From Table \ref{tab:close}, it can be inferred that the designed non-stochastic targeted exploration strategy achieves the exploration goal~\eqref{eq:exp_goal2}. Furthermore, it yields an estimate that is closer to the true estimate compared to the stochastic exploration method.

\begin{table}[t]
\begin{center}
\begin{tabular}{ |c|c|}
\hline 
Exploration& $\|\theta_\mathrm{tr}-\hat{\theta}_T\|$  \\
\hline
Non-stochastic & $0.0257$ \\ 
Stochastic & $0.0389$\\
\hline
\end{tabular}
\caption{Parameter estimation error.}\label{tab:close}
\end{center}
\end{table}

\section{Conclusion}\label{sec:conclusion}
In the proposed work, we design a targeted exploration strategy by using the non-stochastic uncertainty bound based on energy-bounded noise constraints \citep{fogel1979system}. Similar to \citep{venkatasubramanian2023sequential,sarker2023accurate,bombois2021robust}, we consider harmonic/multisine exploration inputs in a priori specified frequencies, since their amplitudes can be easily optimized to yield optimal exploration inputs. As the main contribution, we provide a sufficient condition on the exploration data that guarantees a desired error bound on the estimated parameters. In order to tractably compute the exploration inputs, we neglect the errors in parameters and the effect of disturbances in our proposed strategy, similar to \cite{larsson2016application}. Through a numerical example, we highlight that the profile of the optimal relative energy of the amplitudes for different frequencies of the non-stochastic exploration inputs differs from that of classical stochastic exploration. Furthermore, we demonstrate that the designed non-stochastic targeted exploration strategy results in smaller parametric error, in the presence of unmodeled nonlinearities, compared to classical stochastic exploration.

Achieving targeted exploration without approximations by robustly accounting for noise and parametric uncertainty is part of future work. Similar to~\citep{venkatasubramanian2023sequential}, we expect that the targeted exploration design~\eqref{eq:exp_problem} can be adapted to ensure the resulting model is sufficiently accurate to subsequently design a robust controller.

\bibliography{lit} 
\end{document}